\documentstyle[12pt]{article}
\begin{document}
\newcommand{\first}{1^{\hbox{\em \`ere}}}
\newcommand{\secnd}{2^{\hbox{\em \`eme}}}
\newcommand{\vect}[1]{{\bf #1} }
\newcommand{\bra}[2]{\langle #1,#2 \vert}
\newcommand{\ket}[2]{\vert #1,#2 \rangle}
\newcommand{\vet}[2]{\vert #1,#2)}
\newcommand{\vra}[2]{(#1,#2 \vert}
\newcommand{\cg}[8]{(#1,#2,#3,#4 \vert #5,#6,#7,#8)}
 \newcommand{\qpn}[1]{[[#1]]_{qp}}
 \newcommand{\ti}[1]{\tilde #1}
\newcommand{\somn}{\sum_{n=0}^{\infty}}
\newcommand{\somk}{\sum_{k=0}^{\infty}}
\newcommand{\somp}{\sum_{p=0}^{n}}
\newcommand{\somm}{\sum_{m=0}^{\infty}}


\vskip 4 cm
\centerline {\bf AN $U_{qp}({\rm u}_2)$ ROTOR MODEL FOR ROTATIONAL BANDS}

\centerline {\bf OF SUPERDEFORMED NUCLEI}
\vskip 2 cm
\centerline {R. BARBIER, J. MEYER and M. KIBLER}
\vskip 1 cm
\centerline {Institut de Physique Nucl\'eaire de Lyon,
IN2P3-CNRS et Universit\'e Claude Bernard,}

\centerline {43 Boulevard du 11 Novembre 1918,
             F-69622 Villeurbanne Cedex, France}
\vskip 3 cm

\centerline {\bf Abstract}

\vskip 0.9 true cm

\noindent A nonrigid rotor model is developed from the two-parameter
         quantum algebra $U_{qp}({\rm u}_2)$.
[This model presents the $U_{qp}({\rm u}_2)$ symmetry
and shall be referred to as the $qp$-rotor model.]
A rotational energy formula
as well as a $qp$-deformation
of E2 reduced transition probabilities are derived.
The $qp$-rotor model is applied
(through fitting procedures)
to twenty rotational bands
of superdeformed nuclei in the
 $A \sim 130$, $150$ and $190$ mass regions.
 Systematic comparisons between the $qp$-rotor model and
the $q$-rotor model of
Raychev, Roussev and Smirnov, on one hand, and a basic three-parameter model,
on the other hand,
are performed on energy spectra, on dynamical moments of inertia and
on $B$(E2) values. The physical signification of the deformation
parameters $q$ and $p$ is discussed.

\newpage

\baselineskip = 0.73 true cm

\section{Introduction}
Quantum groups and quantum algebras, introduced at the
beginning of the eightees,$^{1-5}$
continue to attract much attention both in mathematics and
physics.
For the Physicist, a quantum algebra is commonly considered as a deformation
($q$-deformation) of a given Lie algebra.
During the last four years, several works
 have been performed on two-parameter quantum algebras and
quantum groups ($qp$-deformations).$^{6-17}$ (For an elementary
introduction to $q$- and $qp$-quantum algebras, the reader should
consult Ref.~14.)

Most of the physical applications, ranging from chemical
physics to particle physics, have been mainly concerned up to
now with one-parameter quantum algebras ($q$-deformations). In
particular, in nuclear physics we may mention applications to
rotational spectroscopy of deformed and superdeformed nuclei,$^{18-26}$
to the interacting boson model,$^{27,28}$
to the Moszkowski        model,$^{29,30}$
to the U(3) shell        model$ ^{31}$
and to the
Lipkin-Meshkov-Glick     model.$^{32}$ There exist also applications
to particle physics, as for example to quote a few, to hadron
mass formulas$^{33,34}$
and to Veneziano amplitudes.$^{35,36}$ Among
the just mentioned applications, only the ones in
Refs.~25 and 36 rely on the use of
two-parameter deformations.

The aim of the present paper is two-fold: (i) to further
develop and (ii) to apply
(to rotational bands of superdeformed nuclei)
the nonrigid rotor model
briefly introduced in Ref.~25. The latter model,
referred to as the $qp$-rotor model, is based on the
two-parameter quantum algebra $U_{qp}({\rm u}_2)$ while the
$q$-rotor models introduced by Iwao$^{18}$ and
Raychev, Roussev and Smirnov$^{19}$
(see also Refs.~20-23 and 26)
are based
on the one-parameter quantum algebra $U_{q}({\rm su}_2)$.
One of the objectives of this work is to show what we gain when
introducing a second ``quantum algebra''-type parameter, i.e.,
when passing from the
$U_{q}({\rm su}_2)$ symmetry to the $U_{qp}({\rm u}_2)$
symmetry.

The organization of this paper is as follows. The
 $qp$-rotor model is introduced in Sec.~2.
Subsection 2.1 deals with the
mathematical ingredients of the model. The $qp$-rotor model
itself is developed in Subsec.~2.2 (rotational energy formula)
and in Subsec.~2.3 (E2 transition probabilities). Section 3 is
devoted to the application of the $qp$-rotor model to
the description of superdeformed (SD) bands
of nuclei in the $A \sim 130$, $150$ and $190$ mass regions.
The results obtained from the $qp$-rotor model for rotational
energy spectra, dynamical moments of inertia and $B$(E2) values
are compared to the ones derived from the $q$-rotor model
and from a basic ({\em \`a la} Bohr-Mottelson) model. Finally, some
concluding remarks are presented in Sec.~4.

\vskip 2 cm
\section{A $qp$-Rotor Model}

\subsection{The quantum algebra $U_{{qp}}({\rm u}_2)$}
The             quantum algebra $U_{{qp}}({\rm u}_2)$ can be
constructed from two pairs, say $\{ \ti{a}_+^+ ,\ti{a}_+ \}$ and
                                $\{ \ti{a}_-^+ ,\ti{a}_- \}$, of
$qp$-deformed (creation and annihilation) boson
operators. The action of these $qp$-bosons
on a nondeformed two-particle Fock space
$ \{ \ket{n_+}{n_-} : n_+ \in {\bf N} , \
                      n_- \in {\bf N}  \} $
is controlled by
 \begin{eqnarray}
\ti{a}^+_+\;  \ket{n_+}   {n_-} & = &
              \sqrt{ \qpn{n_+ + \frac{1}{2} + \frac{1}{2}}} \;
              \ket{n_+ +1}{n_-} ,   \nonumber \\
\ti{a}_+  \;  \ket{n_+}   {n_-} & = &
              \sqrt{ \qpn{n_+ + \frac{1}{2} - \frac{1}{2}}} \;
              \ket{n_+ -1}{n_-} ,   \nonumber \\
\ti{a}^+_-\;  \ket{n_+}   {n_-} & = &
              \sqrt{ \qpn{n_- + \frac{1}{2} + \frac{1}{2}}} \;
              \ket{n_+}{n_- +1} ,   \nonumber \\
\ti{a}_-  \;  \ket{n_+}   {n_-} & = &
              \sqrt{ \qpn{n_- + \frac{1}{2} - \frac{1}{2}}} \;
              \ket{n_+}{n_- -1} .
\label{eq:qpbo}
\end{eqnarray}
In the present paper, we use the notations
\begin{equation}
\qpn{X} : = { \frac{q^X - p^X}{q-p}}
\label{eq:not1}
\end{equation}
and
\begin{equation}
[X]_q : = [[X]]_{qq^{-1}} = {\frac{q^X - q^{-X}}{q-q^{-1}}} ,
\label{eq:not2}
\end{equation}
where $X$ may stand for an operator or a (real)
number. For Hermitean conjugation requirements, the values of
the parameters
$q$ and $p$ must be restricted to some domains that can be classified
as follows: (i) $q \in {\bf R}$ and
                 $p \in {\bf R}$, (ii)
                 $q \in {\bf C}$ and
                 $p \in {\bf C}$ with $p=q^{\ast}$
(the ${\ast}$ indicates complex conjugation), and (iii)
                 $q = p^{-1} = {\rm e}^{{\rm i} \beta}$ with
$0 \le \beta < 2 \pi$. The two pairs  $\{ \ti{a}_+^+ ,\ti{a}_+ \}$ and
                                        $\{ \ti{a}_-^+ ,\ti{a}_- \}$
of $qp$-bosons commute and satisfy
\begin{equation}
 \ti{a}  _{\pm} \ti{a}^+_{\pm}  =  \qpn{N_{\pm}+1} , \qquad
 \ti{a}^+_{\pm} \ti{a}  _{\pm}  =  \qpn{N_{\pm}  } ,
\label{eq:commu}
\end{equation}
where $N_{+}$ and $N_{-}$ are the usual number operators with
\begin{equation}
N_{\pm} \ \ket{n_+}{n_-} =
n_{\pm} \ \ket{n_+}{n_-} .
\label{eq:openomb}
\end{equation}
Of course, the $qp$-bosons $\ti{a}^+_{\pm}$ and
                           $\ti{a}  _{\pm}$
reduce to ordinary bosons (denoted as
                           $   {a}^+_{\pm}$ and
                           $   {a}  _{\pm}$
in Refs.~37 and 38 and
in Subsec.~2.3) in the limiting situation where $p = q^{-1} \to 1$.

The passage from the (harmonic oscillator) state vectors
$\ket{n_+}{n_-}$          to angular momentum state vectors $\vet{I}{M}$
is achieved through the relations
\begin{equation}
 I := \frac{1}{2}(n_+ + n_-), \quad \qquad
 M := \frac{1}{2}(n_+ - n_-)
\label{eq:sch1}
\end{equation}
and
\begin{equation}
 \vet{I}{M}\  \equiv \ \ket{I+M}{I-M} \  = \  \ket{n_+}{n_-}.
\label{eq:sch2}
\end{equation}
Equations (\ref{eq:qpbo}) may thus be rewritten as
 \begin{eqnarray}
\ti{a}_{\pm}^+ \; \;  \vet{I}{M} & = &
\sqrt{ \qpn{I{\pm}M+\frac{1}{2}+\frac{1}{2}}}\; \;
\vet{I+\frac{1}{2}}{M{\pm} \frac{1}{2}} ,   \nonumber \\
\ti{a}_{\pm}   \; \;  \vet{I}{M} & = &
\sqrt{ \qpn{I{\pm}M+\frac{1}{2}-\frac{1}{2}}}\; \;
\vet{I-\frac{1}{2}}{M{\mp} \frac{1}{2}} ,
\label{eq:qpbo2}
\end{eqnarray}
so that the $qp$-bosons behave as ladder operators for the
quantum numbers $I$ and $M$ (with $\vert M \vert \le I$).

We are now in a position to introduce  a $qp$-deformation of the
Lie algebra ${\rm u}_2$. A simple calculation
shows that the four operators $J_\alpha$ ($\alpha = 0,3,+,-$)
given by
\begin{equation}
J_0 \  := \ {{1}\over{2}}(N_+ + N_-) , \quad
J_3 \  := \ {{1}\over{2}}(N_+ - N_-) , \quad
J_+ \  := \  \ti{a}_+^+ \ti{a}_-     , \quad
J_- \  := \  \ti{a}_-^+ \ti{a}_+
\label{eq:gener1}
\end{equation}
satisfy the following commutation relations$^{14,39}$
 \begin{equation}
[J_3,J_\pm   ] \ = \ \pm J_\pm ,  \qquad
[J_+,J_-     ] \ = \ (qp)^{J_0-J_3} \ \qpn{2J_3} ,  \qquad
[J_0,J_\alpha] \ = \ 0.
\label{eq: comm}
 \end{equation}
We refer to $U_{ {qp}}({\rm u}_2)$ the (quantum) algebra
described by (\ref{eq: comm}). To endow $U_{ {qp}}({\rm u}_2)$ with a Hopf
algebraic
structure, it is necessary to introduce a co-product $\Delta_{qp}$.
The latter co-product is such that:$^{14}$
\begin{eqnarray}
\Delta_{qp}(J_0) & = & J_0 \> \otimes \> 1  +  1 \> \otimes \> J_0, \nonumber
\\
\Delta_{qp}(J_3) & = & J_3 \> \otimes \> 1  +  1 \> \otimes \> J_3, \nonumber
\\
\Delta_{qp}(J_{\pm}) & = & J_{\pm} \> \otimes
\> {(qp)}^{{{1}\over{2}}J_0} \> {(qp^{-1})}^{+{{1}\over{2}}J_3} +
   {(qp)}^{{{1}\over{2}}J_0} \> {(qp^{-1})}^{-{{1}\over{2}}J_3} \>
\otimes \> J_{\pm}
\label{eq:copro}
\end{eqnarray}
and is clearly seen to depend on the two parameters $q$ and
$p$. [Note that with the constraint $p = q^{\ast}$, to be used in
Subsec.~2.2, the co-product
satisfies the Hermitean conjugation property
 $(\Delta_{qp} (J_\pm))^\dagger = \Delta_{pq}(J_\mp)$
and is compatible with the commutation relations for the four operators
$\Delta_{qp} (J_\alpha)$ (with $\alpha = 0,3,+,-$).]
The universal ${\cal{R}}$-matrix (for the coupling
of two angular momenta $I = \frac{1}{2}$) associated to the
co-product $\Delta_{qp}$ reads
\begin{equation}
{\cal{R}}_{pq} = \pmatrix{
p&0&0&0\cr
0&\sqrt{pq}&0&0\cr
0&p-q&\sqrt{pq}&0\cr
0&0&0&p\cr
},
\label{eq:rmatrix}
\end{equation}
and it can be proved that ${\cal{R}}_{pq}$ verifies the so-called
Yang-Baxter equation.

The operator defined by$^{14}$
  \begin{equation}
 C_2( U_{ {qp}}({\rm u}_2))\  :=  \  {{1}\over{2}}(J_+J_- + J_-J_+)\  +\
 {{1}\over{2}} \> \qpn{2} \> (qp)^{J_0-J_3} \>
\left( \qpn{J_3} \right)^2
\label{eq:casim1}
\end{equation}
 is an invariant of the quantum algebra $U_{ {qp}}({\rm u}_2)$. It
 depends truly on the two parameters $q$ and $p$.
The invariant $C_2(U_{{qp}}({\rm u}_2))$ will be one of the main mathematical
ingredients for the $qp$-rotor model to be developed below.
Hence, it is worth to examine its structure more precisely, especially
its dependence on two independent parameters. Equation (\ref{eq:copro})
suggests the following change of parameters
\begin{equation}
Q := (qp^{-1})^{1\over 2} , \qquad
P := (qp)     ^{1\over 2} .
\label{eq:paramet}
\end{equation}
Then, by introducing the generators $A_\alpha$ ($\alpha = 0,3,+,-$)
\begin{equation}
  A_0   \, := \, J_0 ,  \qquad \quad
  A_3   \, := \, J_3 ,  \qquad \quad
  A_\pm \, := \, (qp)^{-{1\over 2}(J_0 - {1\over 2})} \, J_\pm ,
\label{eq:Aoper}
\end{equation}
it can be shown that the two-parameter quantum algebra
 $U_{{qp}}({\rm u}_2)$ is isomorphic to the central
extension
\begin{equation}
U_{qp}({\rm u}_2) = {\rm u}_1 \otimes U_Q({\rm su}_2),
\label{eq:direct}
\end{equation}
where $ {\rm u}_1$ is spanned by the operator $A_0$ and $U_Q({\rm su}_2)$
by the set $\{A_3, A_+, A_-\}$. The $Q$-deformation   $U_Q({\rm su}_2)$
(a one-parameter deformation!)
of the Lie algebra  ${\rm su}_2$ corresponds to the usual commutation
relations
 \begin{equation}
[A_3,A_\pm   ] \ = \ \pm A_\pm ,  \qquad
[A_+,A_-     ] \ = \ [2A_3]_Q  .
\label{eq:acomm}
\end{equation}
Furthermore, the co-product relations (\ref{eq:copro}) leads to
 \begin{equation}
\Delta_{qp}(J_{\pm}) \> = \> P^{ \Delta_{Q}(A_0) - \frac{1}{2} }
\> \Delta_{Q}(A_{\pm}) ,
\label{eq:cpfac}
\end{equation}
where the co-product $\Delta_Q$ is given via
\begin{eqnarray}
\Delta_{Q}(A_0) & = & A_0 \> \otimes \> 1 + 1 \> \otimes \> A_0 , \nonumber \\
\Delta_{Q}(A_3) & = & A_3 \> \otimes \> 1 + 1 \> \otimes \> A_3 , \nonumber \\
\Delta_{Q}(A_{\pm}) & = & A_{\pm} \> \otimes \> Q^{+A_3} +
                                                Q^{-A_3}
                                  \> \otimes \> A_{\pm}.
\label{eq:acopro}
\end{eqnarray}
Equations (\ref{eq:acomm}) involve only one parameter, i.e., the
parameter $Q$. However, two parameters ($Q$ and $P$) occur
in (\ref{eq:cpfac}) as well as in the invariant  $ C_2( U_{ {qp}}({\rm u}_2))$
transcribed in terms of $Q$ and $P$. As a matter of fact,
(\ref{eq:casim1}) can be rewritten as
\begin{equation}
  C_2(U_{qp}({\rm u}_2)) \; = \;
  P^{2A_0-1} \; C_2(U_Q({\rm su}_2)) ,
\label{eq:inv1}
\end{equation}
where
\begin{equation}
C_2(U_Q({\rm su}_2)) \; := \;
  {1 \over 2} \; (A_+A_- + A_-A_+) +
  {1 \over 2} \; [2]_Q \; \left( [A_3]_Q \right)^2
\label{eq:inv2}
\end{equation}
is an invariant of $U_Q({\rm su}_2)$ [compare Eqs.~(\ref{eq:casim1}) and
                                                   (\ref{eq:inv2})].
As a consequence, of central importance for the $qp$-rotor model
of Subsec.~2.2, the invariant $C_2(U_{qp}({\rm u}_2))$, in
either the form (\ref{eq:casim1}) or the form (\ref{eq:inv1}), depends on
two parameters. Finally, it
should be noted that $C_2(U_{qp}({\rm u}_2))$ can be identified
to the invariant of $U_q({\rm su}_2)$ and to the Casimir of $ {\rm su}_2 $
when $p = q^{-1}$ and $p = q^{-1} \to 1$, respectively.
In this sense, the $ U_{ {qp}}({\rm u}_2) $ symmetry
encompasses the
 $ U_{ {q}}({\rm su}_2)$  and  ${\rm su}_2 $ symmetries.

To close this section, let us say a few words on the
representation theory of $U_{{qp}}({\rm u}_2)$ in the case
where neither $q$ nor $p$ are roots of unity. An irreducible
representation of this quantum algebra is described by a Young
pattern
$[\varphi_1; \varphi_2]$ with $\varphi_1 - \varphi_2 = 2I$, where $2I$
is a nonnegative integer ($I$ will represent a spin angular momentum in what
follows). We note
$\vet{[\varphi_1 ; \varphi_2]}{ M }$
(with $M = - I, -I+1,\cdots,+I)$
the basis vectors for the representation
$[\varphi_1 ; \varphi_2]$.

We are now ready to develop a $qp$-rotor model for describing
energy levels and transition probabilities for deformed and
superdeformed nuclei.

 \subsection{Energy levels}
We want to construct a nonrigid rotor model.
As a first basic hypothesis (Hypothesis 1),
we take a rigid rotor with $U_{{qp}}({\rm u}_2)$
symmetry, thus introducing the nonrigidity through
the $qp$-deformation of the Lie algebra $u_2$.
More precisely,
we assume that the $qp$-rotor Hamiltonian $H$
is a
linear function of the invariant $C_2( U_{ {qp}}({\rm u}_2))$:
\begin {equation}
H \; = \; { 1 \over 2{\cal I} } \; C_2 (U_{qp}({\rm u}_2)) + E_0,
\label{eq:ham1}
\end{equation}
where ${\cal I}$  denotes the moment of inertia of the rotor and
$E_0$ the bandhead energy. As a second hypothesis (Hypothesis
2), we
take $\varphi_1 = 2 I$ and
     $\varphi_2 = 0$. This means that we work with
state vectors of the type $\vet{I}{M} \equiv \vet{[2I;0]}{M}$.
Therefore,
the eigenvalues of $H$ are obtained by the action of $H$ on the
physical subspace
$\{ \vet{I}{M} \ : \ M = -I, -I +1, \cdots, +I \}$
of constant angular momentum $I$ corresponding
to the irreducible
representation $[2I;0]$ of $U_{{qp}}({\rm u}_2)$. The two
preceding hypotheses lead to the energy formula
\begin{equation}
  E(I)_{qp}  \; = \; { 1 \over {2 {\cal I}} } \;
  \qpn{I} \; \qpn{I+1} + E_0
\label{eq:eig1}
\end{equation}
for the $qp$-deformed rotational level of angular momentum $I$.

By introducing $s = \ln q$ and $r = \ln p$, Eq.~(\ref{eq:eig1}) yields
\begin{equation}
  E(I)_{qp} \; = \; { 1 \over {2{\cal I}} } \;
      {\rm e}^{ (2I-1) {{s+r} \over 2} } \;
      { {\sinh (I {{s-r}\over 2}) \;
         \sinh  [(I+1) {{s-r}\over 2}]} \over
       {\sinh^2 (  {{s-r}\over 2})} }
      + E_0 .
\label{eq:ener-rs1}
\end{equation}
Preliminary studies have lead us to the conclusion that a good
agreement between theory and experiment cannot be always obtained by
varying  the parameters
$s$ and $r$ (or $q$ and $p$) on the real line ${\bf R}$,
a fact that confirms a similar
conclusion reached in Ref.~20 for $p = q^{-1} \in {\bf R}$. In
addition, if we want that our $qp$-rotor model reduces to the
$q$-rotor model developed by Raychev, Roussev and Smirnov$^{19}$
when $p=q^{-1}$ (or equivalently $r = -s$), we are naturally
left to impose that $(s+r)$ and $(s-r)/{\rm i}$ should be real numbers.
[Observe that the two constraints $(s+r)         \in {\bf R}$
                              and $(s+r)/{\rm i} \in {\bf R}$ ensure
that the energy $E(I)_{qp}$ is real as it should be.]
Furthermore, we shall see that for certain SD bands, a good
agreement between theory and experiment requires that the
parameters $s$ and $r$ vary on the real line ${\bf R}$.
Thus, we shall consider the two possible parametrizations:
\begin{eqnarray}
{\rm (a)} \qquad \qquad
{{s+r}\over 2}         &    =   &  \beta \cos \gamma \in {\bf R}, \qquad
{{s-r}\over 2 {\rm i}} \    =   \  \beta \sin \gamma \in {\bf R}, \nonumber \\
                       &        &                                           \\
{\rm (b)} \qquad \qquad
{{s+r}\over 2}         &    =   &  \beta \cos \gamma \in {\bf R}, \qquad
{{s-r}\over 2 {\rm i}} \    = \ \frac{\beta \sin \gamma}{\rm i}
                                            \in {\rm i} {\bf R}, \nonumber
\label{eq:rs}
\end{eqnarray}
so that the parameters $q$ and $p$ read
 \begin{eqnarray}
{\rm (a)} \qquad \qquad
 q & = &           {\rm e}^{\beta \cos \gamma} \;
          {\rm e}^{+{\rm i} \beta \sin \gamma} , \qquad
 p   =   q^{\ast} \ = \
                   {\rm e}^{\beta \cos \gamma} \;
          {\rm e}^{-{\rm i} \beta \sin \gamma}, \nonumber \\
   &   &                                                  \\
{\rm (b)} \qquad \qquad
 q & = &           {\rm e}^{\beta \cos \gamma} \;
          {\rm e}^{+        \beta \sin \gamma} , \qquad
 p \ = \            {\rm e}^{\beta \cos \gamma} \;
          {\rm e}^{-        \beta \sin \gamma}. \nonumber
\label {eq:qpn1}
\end{eqnarray}
Thus, our $qp$-rotor model involves two independent real
parameters $\beta$ and $\gamma$
corresponding  either to  (a) the two complex
parameters $q$ and $p$ subjected to the constraint
$p = q^{\ast}$     or to  (b) the two real
parameters $q$ and $p$.
In terms of the parameters $\beta$ and $\gamma$,
the  energy formula (\ref{eq:ener-rs1}) takes the
form
\begin {eqnarray}
E(I)_{qq^{\ast}} &  = & { 1 \over {2 {\cal I}} } \;
      {\rm e}^{(2I-1) \beta \cos \gamma} \;
      { {\sin (I \beta \sin \gamma) \; \sin [(I+1) \beta \sin \gamma]} \over
        {\sin^2 (\beta \sin \gamma)} }
      + E_0   \; \; \; \; \quad
              \qquad  \qquad  \quad (27 {\rm a}) \nonumber \\
{\rm or}      \qquad  \qquad  \quad & &          \nonumber \\
E(I)_{qp} \     &  = &  { 1 \over {2 {\cal I}} } \;
      {\rm e}^{(2I-1) \beta \cos \gamma} \;
      { {\sinh (I \beta \sin \gamma) \; \sinh [(I+1) \beta \sin \gamma]} \over
        {\sinh^2 (\beta \sin \gamma)} }
      + E_0   \quad
              \qquad  \qquad  \quad (27 {\rm b}) \nonumber
\label{eq:eig2}
\end{eqnarray}
in the parametrizations of type  (a) or (b), respectively. We shall use
both Eqs.~(27a) and (27b) in our fitting procedures.

\addtocounter{equation}{+1}

In the (a)-parametrization,
to better understand the connection between our $qp$-rotor
model and the $q$-rotor model of Ref.~19, we can perform a
series analysis of Eq.~(27a). A straightforward
calculation allows us to rewrite Eq.~(27a) as
\begin{equation}
E(I)_{qq^{\ast}} \; = \; { 1 \over {2{\cal I}_{\beta \gamma}} } \;
      \left(
\sum_{n=0}^\infty \;
      d_n (\beta,\gamma) \;[I(I+1)]^n +
      (2I + 1) \;
\sum_{n=0}^\infty \;
      c_n (\beta,\gamma) \;[I(I+1)]^n \right) +
      E_0 ,
\label{eq:series}
\end{equation}
 where the expansion coefficients $d_n(\beta, \gamma)$
 and $c_n(\beta, \gamma)$
are given in turn by the series
\begin{eqnarray}
\lefteqn{ d_n(\beta, \gamma)  =
 \frac{2^{2n-1}}{\sin^2(\beta \sin \gamma)} }   \nonumber  \\
  & &  \times  \somk \bigg\{ {(\cos \gamma)}^{2k+2n} \cos(\beta \sin \gamma)
- \cos[(2k+2n  ) \gamma] \bigg\}
 \frac{{\beta}^{2k+2n  }}{(2k+2n  )!}
\frac{(k+n)!}{k! \ n!} ,  \nonumber \\
\lefteqn{c_n(\beta, \gamma)  =   \frac{2^{2n-1}}{\sin^2(\beta \sin \gamma)} }
\nonumber \\
   &  &  \times \somk \  \bigg\{ {(\cos \gamma)}^{2k+2n+1} \cos(\beta \sin
\gamma)
- \cos[(2k+2n+1) \gamma] \bigg\}
 \frac{{\beta}^{2k+2n+1}}{(2k+2n+1)!}
\frac{(k+n)!}{k! \ n!} .
\label{eq:termgene2}
\end{eqnarray}
In Eq.~(\ref{eq:series}), we have introduced the deformed moment of inertia:
\begin{equation}
{\cal I}_{\beta \gamma} = {\cal I} \, {\rm e}^{2 \beta \cos \gamma} ,
\label{eq:inertie1}
\end{equation}
which gives back the ordinary moment of inertia when
$\gamma = \pi/2$ (i.e.,
$q = p^{-1} = {\rm e}^{{\rm i} \beta}$). In the limiting
situation where
$\gamma = \pi/2$, the coefficients
$c_n(\beta, \gamma)$
vanish and the energy formula (\ref{eq:series}) simplifies to
\begin{equation}
E(I)_{qq^{-1}} \; = \; {1 \over 2 {\cal I}} \;
      {{\beta^2} \over {\sin^2 \beta}} \;
      \sum_{n=1}^\infty \; (-1)^{n-1} \;
      {{2^{n-1}} \over {n!}} \;
      \beta^{n-1} \; j_{n-1}(\beta) \; [I(I+1)]^n + E_0,
\label{eq:bonat1}
\end{equation}
where $j_{n-1}$ denotes a spherical
Bessel function of the first kind. Equation (\ref{eq:bonat1})
was derived by
Bonatsos {\em et al}.$^{20}$ for the $q$-rotor model with
$q= {\rm e}^{{\rm i} \beta}$ in order to prove the mathematical
parentage between the $q$-rotor model and the variable moment
of inertia (VMI) model.$^{40-42}$ The
series (\ref{eq:bonat1}) corresponds indeed to the compact expression
\begin{equation}
E(I)_{qq^{-1}} \; = \; {1 \over 2 {\cal I}} \;
      [I]_q [I+1]_q  + E_0 ,
\label{eq:bonat2}
\end{equation}
to be compared with Eq.~(\ref{eq:eig1}).
Note that Eq.~(32) corresponds also to the
(b)-parametrization with $\gamma = \pi / 2$.
The transition from
                    Eq.~(\ref{eq:eig1}) to
                    Eq.~(\ref{eq:bonat2})
illustrates the descent from the $U_{qp}({\rm u}_2)$ symmetry
of the $qp$-rotor to the $U_{q}({\rm su}_2)$ symmetry of the
$q$-rotor. A further descent in
symmetry is obtained when $\beta \to 0$
              (i.e., $q = p^{-1} \to 1$): in this case
$[I]_q [I+1]_q \ \to \  I (I + 1) $ and we get
[from Eq.~(\ref{eq:bonat2})]
the usual energy
formula corresponding to the rigid rotor with ${\rm su}_2$
symmetry.

\subsection{E2 transition probabilities}
 We now examine the implication of the
 $U_{qp}({\rm u}_2)$ symmetry on the calculation of the
 electric quadrupole transition probability $T({\rm E2}; I+2 \to I)$.
 Let us start with the ordinary expression of the reduced
transition probability, namely,
 \begin{equation}
B({\rm E2}; I+2  \to I)  =  \frac {5} {16 \pi}\; Q_0^2\;
 { \bigg  \vert { \cg{I+2}{M}{2}{0}{I+2}{2}{I}{M}} \bigg  \vert} ^2
\label{eq:be2n1}
\end{equation}
for an E2 transition.$^{43}$
In Eq.~(\ref{eq:be2n1}), $Q_0$ is
 the intrinsic electric quadrupole moment in the body-fixed frame.
The coefficient of type $\cg{j}{m}{k}{\mu}{j}{k}{j'}{m'}$ in
the right-hand side of Eq.~(\ref{eq:be2n1}) is a usual
Clebsch-Gordan coefficient for the group SU(2). Our goal is to
find a $qp$-analog of $T({\rm E2}; I+2 \to I)$ and, thus, of
 Eq.~(\ref{eq:be2n1}). The strategy for obtaining a $qp$-analog
of $B({\rm E2}; I+2  \to I)$ is the following:

{(i)} We first rewrite the SU(2) Clebsch-Gordan coefficient of
Eq.~(\ref{eq:be2n1}) in terms of a matrix element of an SU(2)
unit tensor operator $t_{k\mu\alpha}$ with $k=2$, $\mu=0$ and
$\alpha=-2$. This may be done from the    general formula$^{38}$
 \begin{equation}
  \vra{j'}{m'} t_{k \mu \alpha} \vet{j}{m} = \delta(j',j+\alpha)
                                             \delta(m',m+\mu   ) (-1)^{2k}
(2j'+1)^{- \frac{1}{2}} \;  \cg{j}{m}{k}{\mu}{j}{k}{j'}{m'} ,
\label{eq:t202}
\end{equation}
 which shows that the  irreducible  tensor
 operator $t_{k\mu\alpha}$  produces
the  (angular momentum) state vector ${\vet {j+ \alpha}{m+  \mu}}$
  when acting upon the  state vector ${\vet {j}        {m      }}$.
Then, Eq.~(\ref{eq:be2n1}) is amenable to the form
 \begin{equation}
B({\rm E2}; I+2  \to I)    =   \frac {5}{16 \pi}\; {Q}_0^2 \;
 (2I+1 )\;   {\bigg  \vert
\vra{I}{M} t_{20-2} \vet{I+2}{M} \bigg  \vert }^2
\label{eq:be2no1}
\end{equation}
by making use of Eq.~(\ref{eq:t202}).

{(ii)} We know that the general operator  $t_{k\mu\alpha}$
 can be  realized
 in terms of  two pairs  $\{ {a}_+^+ ,{a}_+ \}$ and
                         $\{ {a}_-^+ ,{a}_- \}$
of ordinary boson operators.    In this respect, we
may consider the so-called van der Waerden$^{38}$ realization of
$t_{k\mu\alpha}$.
There are several ways to $qp$-deform the operator $t_{k\mu\alpha}$.
Here, we choose to define a $qp$-deformation $t_{k \mu \alpha}(qp)$
by replacing, in the van der Waerden realization of $t_{k \mu \alpha}$,
the ordinary bosons
$\{ {a}_{\pm}^+ ,{a}_{\pm} \}$ by $qp$-deformed bosons
$\{ \ti{a}_{\pm}^+ ,\ti{a}_{\pm} \}$ and the ordinary
factorials $x!$ by $qp$-deformed factorials
$\qpn{x}! \> = \> \qpn{x  } \>
                  \qpn{x-1} \>
                  \cdots    \>
                  \qpn{1  }$
for $x \in {\bf N}$. We thus obtain
\begin{eqnarray}
 t_{k \mu \alpha}(qp) & = & (-1)^{k+\alpha} \bigg ( {\frac{\qpn{k+\mu}! \;
\qpn{k-\mu}! \; \qpn{k+\alpha}! \;
 \qpn{k-\alpha}! \;  \qpn{2j - k+\alpha}!}{ \qpn{2j + k+\alpha+1}!}}\bigg
)^{\frac{1}{2}} \nonumber \\
                                       &      & \times  \sum_z (-1)^z
{\frac{(\ti{a}_-^+)^{k- \mu -z}
  (\ti{a}_-)^{k - \alpha  -z} (\ti{a}_+^+)^{ \mu + \alpha +z} (\ti{a}_+)^{z}}
{ \qpn{k-\mu -z}! \;  \qpn{k- \alpha - z}! \;  \qpn{\mu + \alpha + z}! \;
\qpn{z}! } } .
\label{eq:opu2}
\end{eqnarray}
In particular, the $qp$-deformed operator $t_{20-2}(qp)$ connecting
the state vector $\vert I+2,M)$, with $j  \equiv I+2$, to
the state vector $\vert I  ,M)$, with $j' \equiv I  $, reads
 \begin{equation}
 t_{20-2}(qp)  =  \bigg ( \frac{\qpn{3} \;
                                \qpn{4} \;
                                \qpn{2I}!}
                               {\qpn{2} \;
                                \qpn{2I +5}!} \bigg )^{\frac{1}{2}}
    (\ti{a}_+)^{2}  (\ti{a}_-)^{2} ,
\label{eq:opu202}
\end{equation}
an expression
of direct interest for deriving the $qp$-analog of $B({\rm E2}; I+2  \to I)$.

{(iii)} We assume that the $qp$-analog
$B({\rm E2}; I+2  \to I)_{qp}$ of $B({\rm E2}; I+2  \to I)$ is simply
 \begin{equation}
B({\rm E2}; I+2  \to I)_{qp}   :=   \frac {5} {16 \pi}\; Q_0^2 \;
   \qpn{2I+1}\;   {\bigg  \vert
\vra{I}{M} t_{20-2}(qp) \vet{I+2}{M} \bigg  \vert }^2 .
\label{eq:qpbe2n2}
\end{equation}
[Equation (\ref{eq:qpbe2n2}) constitutes the third and last
hypothesis (Hypothesis 3) for our $qp$-rotor model.] By using
Eqs.~(\ref{eq:opu202}) and (\ref{eq:qpbo2}),
the relevant matrix element of the
operator $t_{20-2}(qp)$ is easily found to be
  \begin{eqnarray}
\lefteqn{\vra{I}{M} t_{20-2}(qp) \vet{I+2}{M}  = }  \nonumber \\
&  &  \bigg ( \frac{            \qpn{3}     \;
                                \qpn{4}     \;
                                \qpn{2I}!   \;
                                \qpn{I+M+1} \;
                                \qpn{I-M+1} \;
                                \qpn{I+M+2} \;
                                \qpn{I-M+2} }
                                {\qpn{2}    \;
                                \qpn{2I +5}!}  \bigg )^{\frac{1}{2}} .
                                                    \nonumber \\
\label{eq:matx}
\end{eqnarray}
 Then, the introduction of Eq.~(\ref{eq:matx}) into Eq.~(\ref{eq:qpbe2n2})
yields
\begin{equation}
 B({\rm E2}; I+2  \to I)_{qp}    =  { \frac {5} {16 \pi}}\;  Q_0^2 \;
 \frac{\qpn{3} \; \qpn{4}}{\qpn{2}}
\frac{\big(\qpn{I+1} \; \qpn{I+2}\big)^2}{\qpn{2I +2} \; \qpn{2I+3} \;
\qpn{2I+4} \; \qpn{2I+5}}
\label{eq:qpbe2n3}
\end{equation}
in the case of the $K \equiv M = 0$ bands.

For the purpose of comparison with experimental results,  we must calculate
the E2 transition probability $ T( {\rm E2};   I+2  \to I )$
in the $qp$-deformed scheme. We define such a probability by
 \begin{equation}
 T( {\rm E2};   I+2  \to I )_{qp}  :=  1.223 \ 10^9 \
\left( E_{\gamma}(I+2)_{qp} \right)^5  \ B({\rm E2}; I+2  \to I)_{qp}.
\label{eq:taux1}
\end{equation}
Equation (\ref{eq:taux1}) turns out to be a simple
$qp$-deformation of the usual E2 transition probability. [In
     Eq.~(\ref{eq:taux1}), $T({\rm E2}; I+2 \to I)_{qp}$ is in units
     of ${\rm sec}^{-1}$,
 $ E_{\gamma}(I+2)_{qp} := E(I+2)_{qp} - E(I)_{qp}$ in units of
MeV,
and $ B({\rm E2}; I+2  \to I)_{qp} $ in units of  ${\rm e}^2 {\rm fm}^4 $.]

At this stage, a contact with the formula
 $ B({\rm E2}; I+2  \to I)_{q} $ derived by
 Raychev, Roussev and Smirnov$^{19}$
 is in order. First, by taking $p=q^{-1}$ the
right-hand side of (\ref{eq:qpbe2n3}) may be specialized to the
expression of $ B({\rm E2}; I+2  \to I)_{q} $ obtained in Ref.~19.
Hence, our $qp$-rotor model for the E2 transition probability
admits as a particular case the corresponding $q$-rotor
model worked out in Ref.~19. Second, it can be
shown that
 \begin{equation}
  B({\rm E2}; I+2  \to I)_{qp}  =  P^{-4(I+1)}
  B({\rm E2}; I+2  \to I)_{Q },
\label{eq:contact}
\end{equation}
where $P$ and $Q$ are given by (\ref{eq:paramet}).

Let us close with a remark.
Should we have chosen to find a $qp$-analog of the
Clebsch-Gordan coefficient in (\ref{eq:be2n1}), we would have
obtained$^{39}$
 \begin{equation}
{\cg{I+2}{M}{2}{0}{I+2}{2}{I}{M}}_{qp}
=
{\cg{I+2}{M}{2}{0}{I+2}{2}{I}{M}}_{Q }
\label{eq:KAS}
\end{equation}
and, consequently
 \begin{equation}
  B({\rm E2}; I+2  \to I)_{qp}  =
  B({\rm E2}; I+2  \to I)_{Q }.
\label{eq:KASCON}
\end{equation}
We prefer to use (\ref{eq:contact}) rather than
                 (\ref{eq:KASCON})
because the factorization in (\ref{eq:contact}) parallels the
one in (\ref{eq:inv1}).

       \section{Description of Superdeformed Bands}
    \subsection{Fitting procedure}
 The $qp$-rotor model developed in Sec.~2 was applied to twenty
rotational SD bands of
nuclei in the $A \sim 130$, $150$
and $190$ mass regions.
The $\gamma$-ray energies
    \begin{equation}
E_{\gamma}(I) := E(I) - E(I-2)
\label{eq:te}
\end{equation}
were computed from the energy formula
\begin {eqnarray}
 E(I) \; \equiv \; E(I)_{qq^{\ast}} &  = & { 1 \over {2 {\cal I}} } \;
      {\rm e}^{(2I-1) a} \;
      { {\sin (I b) \; \sin [(I+1) b]} \over
        {\sin^2 (b)} }
      + E_0  \qquad \qquad \quad \; \; \; \; \; \; \; \quad
                                          \qquad (46 {\rm a}) \nonumber \\
{\rm or} \qquad \qquad \quad  \qquad  \qquad      & &         \nonumber \\
 E(I) \; \equiv \; E(I)_{qp} \     &  = &  { 1 \over {2 {\cal I}} } \;
      {\rm e}^{(2I-1) a} \;
      { {\sinh (I b) \; \sinh [(I+1) b]} \over
        {\sinh^2 (b)} }
      + E_0  \qquad \qquad \quad \; \; \; \quad
                                          \qquad (46 {\rm b}) \nonumber
\label{eq:eig2p}
\end{eqnarray}
\addtocounter{equation}{+1}
that correspond to Eq.~(27a) or (27b), respectively, with
    \begin{equation}
a \; := \;  \beta \cos \gamma, \qquad \quad
b \; := \;  \beta \sin \gamma.
\label{eq:abbg}
\end{equation}
The free parameters of the $qp$-rotor model are then
$a$, $b$ and ${\cal I}$.

For the sake of comparison, we also computed the transition
energies $E_{\gamma}(I)$ and performed an analysis of the
same SD bands from two other models. First, we used
   \begin{equation}
E(I) \; \equiv \; E(I)_{qq^{-1}} \; = \; { 1 \over {2 {\cal I}'} } \;
            { {\sin (I \beta' ) \; \sin[ (I+1) \beta'] } \over
        {\sin^2 \beta' } } + E_0,
\label{eq:eigq}
\end{equation}
 where $\beta'$ (defined by $q = {\rm e}^{{\rm i} \beta'}$)
and ${\cal I}'$ are the two free parameters of the $q$-rotor
model of Ref.~19. Second, we also applied the energy formula
    \begin{equation}
E(I) \; \equiv \; A'I(I+1) \; + \; B' [I(I+1)]^2
                           \; + \; C'[I(I+1)]^3
                           \; + \;  E_0
\label{eq:eig3}
\end{equation}
arising from the (Bohr-Mottelson)
basic model$^{43}$ restricted to three free parameters,
i.e., $A'$, $B'$ and $C'$.

For each of the three models, the parameters
($a$, $b$  and ${\cal I}$ for the $qp$-rotor;
 $\beta'$  and ${\cal I}'$ for the $q$-rotor;
$A'$, $B'$ and $C'$ for the basic model)
were fixed by minimizing
    \begin{equation}
\chi  := \sqrt{ { 1 \over {n -  m}} \; \sum_I \;
         \left[ { {E_{\gamma}^{\rm th } (I) -
                   E_{\gamma}^{\rm ex } (I)} \over
           {\Delta E_{\gamma} (I)} } \right]^2},
\label{eq:qui2}
\end{equation}
where $n$
is the number of experimental points included in the fitting
procedure, $m$ is the number of freely varied parameters, and
 $\Delta E_{\gamma} (I)$ are the experimental errors.

 \subsection{Results and discussions}
 \subsubsection{Fitting of data}
We present in Table 1  the free parameters and the ${\chi}$-values,
for the $q$- and the $qp$-rotor models, obtained from the twenty
fitted SD bands.
(The tables and figures of this paper
can be obtained from the authors.)
For space saving purposes, we do not report the corresponding
results obtained with the basic model since the ${\chi}$-values
are generally higher than the ones
derived from
the $qp$-rotor model (a fact to be confirmed in Subsec.~3.2.2).

Table 1 exhibits two general trends:
first,  the best results are obtained in the $ A \sim 190 $ mass region
for
the two models   and, second,
the  ${\chi}$-values for the
$qp$-rotor are better than those for the $q$-rotor.
 Indeed, the $\chi$-values obtained for the $qp$-rotor
(respectively, $q$-rotor)
are between 0.6 and 4.2 (respectively, 0.9 and 29.4)
in the  $A \sim 190$ mass region except for
 $^{194}{\rm Hg(a)} $ with $\chi = 9.1$ (respectively, 35.2)
while the $\chi$-values are between 1.9 and 20.3
(respectively, 8.8 and 87.9) in the two other mass regions.
The high values obtained for $\chi$ are not surprising:
in the standard definition of $\chi$, Eq.~(\ref{eq:qui2}),
 the difference between the theoretical and experimental transition energies
is divided
by the experimental error $\Delta E_{\gamma} (I)$ that
is equal to 0.5 keV except for the recent experimental
data$^{44,45}$ on
$^{192}{\rm Hg   } $,
$^{194}{\rm Hg(a)} $ and
$^{194}{\rm Hg(b)} $
for which
$\Delta E_{\gamma} (I)$ is as below as 0.1 keV.
Therefore,   we may emphasize
the excellent quality of the fits for the $A \sim 190$ bands,
especially in the case of the $qp$-rotor model.

For the $qp$-rotor model, the quality of the fits is connected
to the nature (real or complex) of the parameters $q$ and $p$.
The best fits were obtained by taking:
(i)  the (a)-parametrization (i.e.,
$q = {\rm e}^{a + {\rm i} b}$ and
$p = {\rm e}^{a - {\rm i} b}$) for $^{146}$Gd and the 190 SD bands
and
(ii) the (b)-parametrization (i.e.,
$q = {\rm e}^{a + b}$ and
$p = {\rm e}^{a - b}$) for the 130 and 150 SD bands.
To illustrate our results, we globally
characterize in Fig.~1 the twenty
SD bands by their position in the plane of the two
``quantum algebra''-type real parameters $a$ and $b$.
As it was shown in Ref.~20, the parameter
$\beta '$ of the $q$-rotor
(which occurs in a sine like the parameter $b$
in the (a)-parametrization of the $qp$-rotor)
can be interpreted as a softness or stretching parameter of the nucleus,
similar to the parameter $\sigma$ of the VMI model.$^{40-42}$
We adopt this interpretation for the parameter $b$
(that coincides with the parameter $\beta'$ when $a=0$)
in the (a)-parametrization.
Then, the (b)-parametrization describes a distortion phenomenon
(decrease of the dynamical moment of inertia with the spin of the nucleus)
rather than a stretching phenomenon.
In the (a)- and (b)-parametrizations, the role played by the
parameter $a$, appearing in the exponential term of
Eq.~(\ref{eq:eig2p}), is clear. The parameter $a$ has a crucial
effect of correction on the distortion (stretching or anti-stretching)
function of the  parameter $b$.
In addition, we note from Table~1 that the sign of $a$
is the same as that of the difference
${\cal I} - {\cal I}'$.
In other words, at high angular
momenta, the exponential term in
(\ref{eq:eig2p}) moderates (when $a<0$) or accentuates (when $a>0$)
the contribution to the energy
of $\frac{1}{2{\cal I}}$ with respect to the contribution
of $\frac{1}{2{\cal I}'}$.
Therefore, the parameter $a$
can moderate or accentuate the distortion phenomenon
of the nucleus.

Before performing a systematic comparison between the three models
under consideration,
 we present in Tables 2-6
 the calculated and experimental  transition energies for the $qp$-rotor model.
The numerical results in Tables 2-6 confirm the preceding
interpretation of the
free parameters of the $qp$-rotor model.
Here again, we note that the quality of the fits is better in the
$A \sim 190$ region than in the $A \sim 130$ and $150$
regions. This reflects the fact that the $\gamma$-ray energies
range from 200 to 900  keV (respectively,
           600 to 1700 keV) for angular momenta ranging
from 8 to 50 (respectively, 14 to 64) for
$A \sim 190$ (respectively, $A \sim 130$ and $150$).

\subsubsection{Comparative analyses}
In order to confirm the difference (already evocated in
the $\chi$-values analysis) between the $qp$-rotor
model and the basic model, we consider three representative
nuclei for each of the considered mass regions.
Figure 2  shows the differences between the
calculated and experimental transition energies
for the nuclei $^{132}{\rm Ce}$,
               $^{152}{\rm Dy}$ and
               $^{192}{\rm Hg}$
obtained from the basic and $qp$-rotor models.
It is clear that the $qp$-rotor model is more appropriate,
in particular  for  $^{132}$Ce,
for describing the distortion phenomenon
than the basic model.
Therefore, we switch to a detailed comparison between the
$q$- and $qp$-rotor models.
Figures 3-7
display the
results (in terms of differences as in Fig.~2)
afforded by the $q$- and $qp$-rotor models
for the twenty SD bands under study. Two remarks arise from Figs.~3-7.
First, the  preceding ${\chi}$-values analysis is clearly confirmed.
Second, we observe that the $qp$-rotor model is much
better than the $q$-rotor one when
the distortion phenomenon is particularly pronounced.
For example, in the case
of the $^{192}{\rm Hg}$ band that presents nineteen transitions
and
where the variation of the stretching effect becomes less important at
high spin,
the $qp$-rotor model provides the best results.

An alternative way to analyse the stretching phenomenon
in the $A \sim 190$ region amounts to compare the {\em theoretical}
and {\em experimental} dynamical moments of inertia
\begin{equation}
{\cal I}_{{\rm th}}^{(2)}(I) :=
\left( \frac { d ^2 E } { d x ^2 }  \right)^{-1},
\quad E \equiv E(I),
\quad x \equiv x(I) := \sqrt{I(I+1)}
\label{eq:inertie2}
\end{equation}
and
\begin{equation}
{\cal I}_{{\rm ex}}^{(2)}(I) := \frac{4000 } {E_{\gamma} (I+2)-E_{\gamma}(I) },
\label{eq:inertie3}
\end{equation}
respectively.
The  experimental $\gamma$-ray energies $E_{\gamma}$
in (\ref{eq:inertie3}) are defined by
 (\ref{eq:te}) and we take the theoretical energies
$E$ in (\ref{eq:inertie2}) as
given by (\ref{eq:eig2p}) (respectively,
(\ref{eq:eigq})) for the
$qp$-rotor (respectively, $q$-rotor) model.
[The dynamical
moments in (\ref{eq:inertie2}) and \ref{eq:inertie3})
are in units of $\hbar^2 {\rm MeV}^{-1}$.]
Figure 8 shows the results for four SD bands of the three nuclei
$^{190-192-194}{\rm Hg}$: the experimental moments of inertia are
calculated from Refs.~44, 45 and 54
and the theoretical ones by using the free parameters
${ 1 \over {2 {\cal I} } }$, $a$, $b$ and
${ 1 \over {2 {\cal I}'} }$, $\beta '$
for the  $ U_{ qp}({\rm u }_2)$  and
         $ U_{ q }({\rm su}_2)$ symmetries,
respectively.
The $qp$-rotor results are much
closer to the experimental results than the
$q$-rotor ones, due to the
influence of the parameter $a$. [In passing,
Fig.~8 shows that globally, both for the $q$- and $qp$-rotor models,
the second derivative of the
energy is significative when calculated
with the fitted values of the free parameters.]

A last way to compare the $qp$-rotor model with the two others
is to use experimental values of E2 transition probabilities.
{}From such values, we can compute two different
intrinsic electric quadrupole moments, namely,
$(Q_0)_{qp}$ and
$(Q_0)_{q }$ for the
$ U_{ qp}({\rm  u}_2)$ and
$ U_{ q }({\rm su}_2)$ symmetries, respectively.
For the $ U_{ qp}({\rm  u}_2)$ symmetry,
$(Q_0)_{qp}$ is deduced from
(\ref{eq:qpbe2n3})
and
(\ref{eq:taux1}),
where we take the experimental value for the E2 transition
probability and all the other terms (including the transition
energies) are calculated from the parameters of the $qp$-rotor
model obtained from the optimization of energy. A similar
calculation is conducted for
$(Q_0)_q$ corresponding to the $ U_{ q }({\rm su}_2)$ symmetry.
The {\em experimental} intrinsic electric quadrupole moment
$(Q_0)_{\rm ex}$ corresponds to the su$_2$ symmetry: it is
calculated from (\ref{eq:qpbe2n3})
and (\ref{eq:taux1}) with $q= p^{-1} \to 1$ by taking
the experimental E2           transition probability and
the experimental $\gamma$-ray                 energy.
 We present in Tables 7-9  the values of the quadrupole moments
 $(Q_0)_{qp    }$,
 $(Q_0)_{q     }$ and
 $(Q_0)_{\rm ex}$, together with the experimental errors,
computed
  for $^{192    }{\rm Hg}$ and
      $^{194-196}{\rm Pb}$
with the experimental E2 transition probabilities of Refs.~57, 60
and 61. We see that the values of $(Q_0)_{qp}$ are in better agreement
with the experimental quadrupole moments
 $(Q_0)_{\rm ex}$ than the values of $(Q_0)_{q}$.
To further compare the symmetries
$ U_{ qp}({\rm u }_2)$,
$ U_{ q }({\rm su}_2)$ and
${\rm su}_2$, it is interesting to calculate the geometrical
factor of type
 \begin{equation}
 G(I) := \frac{ 16 \pi } {5} \frac {B({\rm E2}; I+2 \to I)} {5Q_0^2}
 \label{eq:finale}
 \end{equation}
for the three symmetries.
Figure 9 displays this factor as a function of the spin $I$
for the three nuclei $^{192    }{\rm Hg}$ and
                     $^{194-196}{\rm Pb}$.
At high spin, the increasing of $G(I)$ characterises the two
``quantum algebra''-type models, while $G(I)$ reaches a limit
value for the su$_2$ symmetry.
Note that $G(I)$ increases less strongly
(i.e., more linearly)
for the $U_{qp}({\rm  u}_2)$ symmetry than
for the $U_{q }({\rm su}_2)$ symmetry
when the parameter $b$ differs from the parameter $\beta '$.

\section{Conclusions}
In this paper, we concentrated
on a new nonrigid rotor model (the $qp$-rotor model)
based on three hypotheses in the framework
of an investigation of the two-parameter quantum
algebra $U_{ qp}({\rm u}_2)$.
The two facets of this model
consist of a
 three-parameter energy level formula
and
a $qp$-deformed E2 transition probability formula.
As limiting cases, the $qp$-rotor model gives back
the $q$-rotor model$^{19}$ (when $p = q^{-1}$)
based on the quantum algebra $U_{ q}({\rm su}_2)$ and
the rigid rotor model (when $p = q^{-1} \to 1$)
based on the Lie algebra su$_2$.

Twenty rotational bands of
superdeformed nuclei in the  $A \sim  130$, $150$ and $190$ mass regions
were used to test
our $qp$-rotor model
and to compare it to the $q$-rotor model and to a basic (with a three-term
polynomial energy formula) model.
The main results may be summarized as follows.
First, the $qp$-rotor model is better than
       the $q $-rotor model and the
                basic model as far as energy spectra are concerned.
 Second, the energy fits for the twenty SD bands
are in good agreement with
experiment both for the $q$- and $qp$-rotor models.
However, a marked difference between the
latter two models manifests itself in the energy spectrum and also
in the second derivative of the energy
(i.e., for the dynamical moment of inertia).
Third, in terms of  $B$(E2)   values the results afforded by the
 $U_{ qp}({\rm  u}_2)$ symmetry are between those given by the
 $U_{ q }({\rm su}_2)$ symmetry and the su$_2$ symmetry: the
 $B$(E2)   values for the $qp$-rotor model increase more or less linearly with
spin, a result that does not hold for the  $q$-rotor model.

As a general conclusion, the $qp$-rotor is appropriate for
describing the collective phenomenon of distortion
occurring in the rotation of the nucleus
(increase or decrease of the dynamical moment of inertia with the spin).
The net difference
between the $q$- and $qp$-rotor models comes from the ``quantum
algebra''-type parameter $a$ that tends to smooth the
(spherical or hyperbolical) sine term in the energy
and thus accentuates or moderates the distortion phenomenon of the nucleus.

To close this paper, let us mention that Hypothesis 2 (i.e.,
$\varphi_1 = 2I$ and $\varphi_2=0$)
of our model might be abandoned. This would lead to a {\em \`a la}
Dunham formulation
for describing
more complicated rotational spectra of
deformed and superdeformed nuclei or
rovibrational spectra of diatomic molecules.
As a further extension, it would be also interesting to combine our model with
one of Ref.~24 (based on the $q$-Poincar\'e symmetry) in the case of heavy
nuclei. Work in these directions is in progress.

\vskip 1 cm

\noindent {\bf Acknowledgments}

\noindent The authors would like to thank M.~Meyer, N.~Redon
and Yu.~F. Smirnov for valuable discussions. They are also
indebted to M. Tarlini for calling their attention to
Ref.~24. Part of this work
was presented by one of the authors (M.~K.)
to the international symposium ``Symmetries in
Science, VIII'' organized by B. Gruber (Bregenz, Austria,
August 1994). In this connection, thanks are due to
V.~K. Dobrev, H.-D. Doebner,
F. Iachello and S. Nishiyama for pertinent
remarks.

\newpage

\noindent {\bf Table captions:}
\vskip 0.25 cm

\noindent
Table 1. Free parameters for the $q$- and $qp$-rotor models:
$\beta'$
corresponds to $q= {\rm e}^{{\rm i} \beta'}$;
$a =\beta\cos\gamma$ and
$b =\beta\sin\gamma$
correspond: (i) to
$ q  =  {\rm e}^{a + {\rm i}b } $ and
$ p  =  {\rm e}^{a - {\rm i}b } $
for $^{146}$Gd and the 190 SD bands, and (ii) to
$ q  =  {\rm e}^{a + b } $ and
$ p  =  {\rm e}^{a - b } $
for the 130 and 150 SD bands;
${1\over{2{\cal I} }}$ and
${1\over{2{\cal I}'}}$
are  in units of  $\hbar^{-2}$keV.
\vskip 0.25cm

\noindent
Table 2. Theoretical and experimental $\gamma$-ray  energies
and experimental  errors for SD bands in the $ A \sim 130 $
region. Experimental data  are taken from
Refs.~46-48.
\vskip 0.25cm

\noindent
Table 3. Theoretical and experimental $\gamma$-ray  energies
and experimental  errors for SD bands in the $ A \sim 150 $
region. Experimental data  are taken from
Refs.~49-53.
\vskip 0.25cm

\noindent
Table 4. Theoretical and experimental $\gamma$-ray  energies
and experimental  errors for SD bands in the $ A \sim 190 $
region. Experimental data  are taken from
Refs.~44, 45 and 54.
\vskip 0.25cm

\noindent
Table 5. Theoretical and experimental $\gamma$-ray  energies
and experimental  errors for SD bands in the $ A \sim 190 $
region. Experimental data  are taken from
Refs.~55-58.
\vskip 0.25cm

\noindent
Table 6. Theoretical and experimental $\gamma$-ray  energies
and experimental  errors for SD bands in the $ A \sim 190 $
region. Experimental data  are taken from
Ref.~59.
\vskip 0.25cm

\noindent
Table 7. Intrinsic electric
quadrupole moments for  $^{192}$Hg
in units of  eb.
The theoretical moments $(Q_0)_{q }$ and
                        $(Q_0)_{qp}$
are  calculated with
${1\over{2{\cal I}'}} = 5.58 $,
$\beta' = 0.12\; 10^{-1} $
 and
${1\over{2{\cal I }}} = 5.91  $,
$a = -0.15\; 10^{-2}$,
$b =  0.47\; 10^{-2}$
for the $q$- and $qp$-rotor models,
respectively.
The experimental  values   $(Q_0)_{\rm ex}$
as well as the upper and lower experimental errors
${\Delta}Q_0^{+} $ and
${\Delta}Q_0^{-} $
are taken from Ref.~60.
\vskip 0.25cm

\noindent
Table 8. Intrinsic electric
quadrupole moments for  $^{194}$Pb
in units of  eb.
The theoretical moments $(Q_0)_{q }$ and
                        $(Q_0)_{qp}$
are  calculated with
${1\over{2{\cal I}'}} = 5.62 $,
$\beta' = 0.13\; 10^{-1} $
 and
${1\over{2{\cal I }}} = 5.75  $,
$a = -0.78\; 10^{-3}$,
$b =  0.92\; 10^{-2}$
for the $q$- and $qp$-rotor models,
respectively.
The experimental   values           $(Q_0)_{\rm ex}$
as well as the upper and lower experimental errors
${\Delta}Q_0^{+} $ and
${\Delta}Q_0^{-} $
are taken from Ref.~61.
\vskip 0.25cm

\noindent
Table 9. Intrinsic electric
quadrupole moments for  $^{196}$Pb
in units of  eb.
The theoretical moments $(Q_0)_{q }$ and
                        $(Q_0)_{qp}$
are  calculated with
${1\over{2{\cal I}'}} = 5.68 $,
$\beta' = 0.11\; 10^{-1} $
 and
${1\over{2{\cal I }}} = 5.71  $,
$a = -0.17\; 10^{-3}$,
$b =  0.11\; 10^{-1} $
for the $q$- and $qp$-rotor models,
respectively.
The experimental   values             $(Q_0)_{\rm ex}$
as well as the upper and lower experimental errors
${\Delta}Q_0^{+} $ and
${\Delta}Q_0^{-} $
are taken from Ref.~57.

\newpage

\noindent {\bf Figure captions:}
\vskip 0.5cm

\noindent
Fig. 1.
The characterization, in the plane of the free parameters
     $a =\beta\cos\gamma$ and
     $b =\beta\sin\gamma$
of the $qp$-rotor model, of the SD bands in the
$ A \sim 130 $, 150 and 190 mass regions.
\vskip 0.5cm

\noindent
Fig. 2. Comparison between the theoretical
and experimental  $\gamma$-ray  energies in keV.
Solid lines and dotted lines display the results for
the $qp$-rotor model and the basic  model,
 respectively.
\vskip 0.5cm

\noindent
Fig. 3. Comparison between the theoretical
and experimental  $\gamma$-ray  energies in keV.
Solid lines and dotted lines display the results for
the $qp$-rotor model and the $q$-rotor  model,
 respectively.
\vskip 0.5cm

\noindent
Fig. 4. Comparison between the theoretical
and experimental  $\gamma$-ray  energies in keV.
Solid lines and dotted lines display the results for
the $qp$-rotor model and the $q$-rotor  model,
 respectively.
\vskip 0.5cm

\noindent
Fig. 5. Comparison between the theoretical
and experimental  $\gamma$-ray  energies in keV.
Solid lines and dotted lines display the results for
the $qp$-rotor model and the $q$-rotor  model,
 respectively.
\vskip 0.5cm

\noindent
Fig. 6. Comparison between the theoretical
and experimental  $\gamma$-ray  energies in keV.
Solid lines and dotted lines display the results for
the $qp$-rotor model and the $q$-rotor  model,
 respectively.
\vskip 0.5cm

\noindent
Fig. 7. Comparison between the theoretical
and experimental  $\gamma$-ray  energies in keV.
Solid lines and dotted lines display the results for
the $qp$-rotor model and the $q$-rotor  model,
 respectively.
\vskip 0.5cm

\noindent
Fig. 8. The dynamical moments of inertia for
$^{190-192}$Hg,
$^{194    }$Hg(a) and
$^{194    }$Hg(b)
calculated for the
$U_ q  ({\rm su}_2)$  and
$U_{qp}({\rm  u}_2)$ symmetries and
compared to the experimental values.
The moments of inertia are in units of
$\hbar^{2}$MeV$^{-1}$.

\vskip 0.5cm
\noindent
Fig. 9. The geometrical factor $G(I)$
of the reduced transition probabilitiy $B$(E2) for
$^{192}$Hg,
$^{194}$Pb and
$^{196}$Pb
calculated for the
$U_q   ({\rm su}_2)$,
$U_{qp}({\rm  u}_2)$ and
$       {\rm su}_2$  symmetries.

\end{document}